\documentstyle[12pt]{article}
\newcommand{\be}{\begin{eqnarray}}
\newcommand{\ee}{\end{eqnarray}}

\begin{document}

\begin{center}
{\LARGE\bf Relativistic Ring-Diagram Nuclear Matter Calculations}
\\
\vspace*{.5cm}
{\large\sc M.F. Jiang$^a$, R. Machleidt$^b$, and T.T.S. Kuo$^c$}
\\
{\it $^a$Department of Physics \& Astronomy,
Vanderbilt University,
Nashville, Tennessee 37235 \\
$^b$Department of Physics, University of Idaho,
Moscow, Idaho 83843 \\
$^c$Department of Physics,
State University of New York at Stony Brook,
Stony Brook, New York 11794}
\\
\today
\end{center}
\vspace*{.5cm}

\begin{abstract}
A relativistic extension of the
particle-particle hole-hole
ring-diagram many-body formalism is developed
by using the Dirac equation for single-particle motion in the medium.
Applying this new formalism, calculations are performed for nuclear matter.
The results show that the saturation density is improved and the
equation of state becomes softer as compared to corresponding
Dirac-Brueckner-Hartree-Fock calculations.
Using the Bonn A potential,  nuclear
matter is predicted to saturate at
an energy per nucleon of --15.30 MeV
and a density equivalent to a Fermi momentum of 1.38 fm$^{-1}$,
in excellent agreement with empirical information.
The compression modulus is 152 MeV at the saturation point.
\end{abstract}

\section{Introduction}

The microscopic explanation of the
properties of nuclear matter is one of the
most fundamental traditional issues in theoretical nuclear physics. Being an
infinite and homogeneous nucleon assembly, nuclear matter not only serves as
an ideal system to test many-body theories but also provides
constraints on the off-shell part of the
nucleon-nucleon (NN) interaction which can not be
determined by studying two-nucleon scattering data
and the properties of the deuteron.

For many years,
conventional Brueckner
theory and variational methods have been widely used in the study of nuclear
matter. However, there appears to be an intrinsic problem
associated with these approaches. All calculations using these methods,
no matter which potential is applied, fail to
reproduce the empirical saturation properties of nuclear matter
(energy per nucleon $E/A\approx -16$ MeV and saturation density
$\rho_0\approx 0.17$ fm$^{-3}$)~\cite{Day,DW85,Mac}.
To improve conventional nuclear matter
predictions,  two types of
additional effects are needed:
 a general increase of the binding energy and a strongly
density-dependent repulsive contribution.

In the early 1980's, the Brooklyn group~\cite{Ana83}
 motivated and worked out a relativistic extension of
Brueckner theory, which was further developed by Horowitz and
Serot~\cite{HS84}, by Brockmann and Machleidt~\cite{BM84,BM90},
 and by ter Haar and
Malfliet~\cite{TM87}.
The basic idea of these so-called Dirac-Brueckner approaches is the use of the
Dirac equation for the single particle motion in nuclear matter.
Due to the
density dependence of the effective nucleon mass in the medium,
the Dirac-spinor wave functions representing the nucleons become density
dependent, and so does the nucleon-nucleon (NN) interaction. As a consequence
of this, the  attractive $\sigma$-boson
exchange is suppressed causing a repulsive effect, which strongly increases
with density. This effect allows to explain the empirical saturation density of
nuclear matter. The correct binding energy of nuclear matter is
obtained if for the single particle (s.p.) potential
a continuous choice is applied and if a weak tensor-force NN interaction is
used; both these factors increase the binding energy~\cite{Mac,BM90}.

So far, the Dirac-Brueckner
approach has been applied only in the lowest order in
the reaction matrix, $K$ [Dirac-Brueckner-Hartree-Fock (DBHF)].
 For non-relativistic Brueckner theory it is
known that contributions beyond the lowest order in $K$ are not necessarily
small. Therefore, it is most desirable to also investigate for the
Dirac-Brueckner approach contributions of higher orders. It is the purpose of
this paper to start such an investigation. As a first class of diagrams beyond
past efforts, we will consider the ring-diagram series displayed
in Fig.~1. In the non-relativistic many-body frame work, these
 diagrams have been investigated by the Stony Brook
group~\cite{YHK,SYK,JKM,JHYK,JYK,JMK90}.

In the Stony Brook ring approach, hole-hole (hh) propagations in
intermediate
states are taken into account. These long-range correlations are ignored in
standard Brueckner theory for nuclear matter.
 Short-range correlations [particle-particle (pp)
propagations] are included in the ring formalism, as they are in Brueckner
theory. In the non-relativistic calculations, it is found that the hole-hole
propagations provide an effect that is repulsive and strongly
density-dependent~\cite{JMK90a,Ramos}. Thus, nuclear matter saturation is
improved. Moreover, the nuclear matter equation-of-state becomes softer, i.~e.,
the compression modulus is reduced.

As a matter of fact, both the relativistic effects as obtained in  DBHF
and the hh
ring contributions shift the saturation density towards more realistic values.
Yet, the predicted saturation densities are still above the empirical value.
In the non-relativistic
ring approach, one gets typically $k_F
\raisebox{-.3ex}{\small $ \stackrel{\textstyle >}{\sim}$}
 1.42$ fm$^{-1}$, and in
DBHF
calculations, $k_F
\raisebox{-.3ex}{\small $ \stackrel{\textstyle >}{\sim}$}
 1.40$ fm$^{-1}$ at the saturation point.
 Thus, there is still room for improvement.
Therefore, it will be interesting to see what result the
 coherent contributions will yield. In other words,
 we like to investigate what role the ring-diagram effect will play
in a relativistic framework, and whether in this approach
 the  saturation density will be lowered even more than in the DBHF
calculations. By doing so, we can at least show partial corrections to the
DBHF results from certain diagrams of higher order in the reaction matrix.

In Section 2 we will explain the Dirac approach and its application to the ring
formalism. Results are presented and discussed in Section 3. Finally, Section~4
contains a summary and some conclusions.

\section{The Relativistic Ring-Diagram Approach}

As mentioned, the essential point of the relativistic many-body approach is
the use of the Dirac equation to describe the single-particle
motion in the nuclear medium (notation as in the textbook
by J.D. Bjorken and S.D. Drell~\cite{BD})
\be
(\not p-M-\Sigma)\tilde u({\bf p}, s)=0
\ee
with the relativistic self-energy operator
\be
\Sigma=U_s+\gamma_0U_v
\ee
where ${U_s}$ is an attractive scalar and ${U_v}$ the time-like component of a
repulsive vector field. These two fields
are strongly density-dependent. The momentum dependence is weak and will
 therefore
be neglected.

The solution of Eq.(1) is
\be
\tilde u({\bf p}, s)=\sqrt{\frac{\tilde E_p+\tilde M}{2\tilde E_p}}
\left(\begin{array}{c}1 \\
\frac{\mbox{\boldmath$\sigma$}\cdot {\bf p}}{\tilde E_p+\tilde M}
\end{array} \right)\chi_s
\ee
with
\be
\tilde M=M_N+U_s\hskip 1cm \mbox{\rm and}\hskip 1cm
\tilde E_p=\sqrt{\tilde M^2+{\bf p}^2},
\ee
and ${\chi_s}$ a Pauli spinor. The normalization is ${\tilde
u^{\dag}\tilde u=1}$.

In the approximation of an instant NN interaction, we can use
two-time Green functions to define the particle-particle hole-hole propagations
in the relativistic framework.
Neglecting negative-energy intermediate states
(i.e. anti-nucleon contributions), the free two-particle
propagator, in the basis defined by the self-consistent Dirac s.p.
wavefunctions, can be written as
\be
\tilde F^{pphh}_{\alpha\beta\gamma\delta}(12, \omega)
=\{\frac{\bar n_1\bar n_2}{\omega-(\epsilon_1+\epsilon_2)+i0}
-\frac{n_1n_2}{\omega-(\epsilon_1+\epsilon_2)-i0}\}
(\Lambda_1)_{\alpha\gamma}(\Lambda_2)_{\beta\delta}
\ee
where ${\epsilon_i=\tilde E_i+U_v}$ is the relativistic s.p. energy, and
\be
\Lambda_i=\frac{\tilde{\not p}_i+\tilde M}{2\tilde M}
=\frac{\tilde E_i\gamma_0-{\bf p}_i.{\mbox{\boldmath$\gamma$}}+\tilde
M}{2\tilde M}. \ee
is a positive-energy nucleon projection operator.

With this propagator, the series of pphh ring diagrams shown in
Fig. 1 can be summed up similarly to the non-relativistic
case~\cite{SYK,JKM}.
Summing up
this series of infinite ring diagrams, the ground-state energy shift of nuclear
matter is given by
\be
\Delta E^{pphh}_0=\frac{-1}{2\pi i}\int^{\infty}_{-\infty}d\omega
e^{-i\omega_0^+}
Tr_P\{\tilde F\tilde K+\frac{1}{2}(\tilde F\tilde K)^2+\frac{1}{3}(\tilde
F\tilde K)^3+\ldots\}
\ee
where ${\tilde K}$ stands for a relativistic model-space Dirac-Brueckner
reaction matrix (see below).  ${Tr_P}$ represents a
trace operation within the model-space $(P)$, i.e a sum over all states
with ${|{\bf p}|\leq
k_{_M}}$. Here ${k_{_M}}$ is the momentum space boundary of the model-space.
In this study, we use for ${k_{_M}}$ the same value as in our previous
work\cite{SYK,JKM}, namely ${k_{_M}}$=3.2 fm$^{-1}$.

To calculate the relativistic reaction matrix $\tilde K$,
we employ the relativistic three-dimensional
 Thompson equation~\cite{Tho71}
in nuclear matter~\cite{BM90,MMB90}
\be
\tilde K_{ijkl}(\omega)=\frac{1}{2}\tilde V_{ijkl}
+\frac{1}{2}\sum_{mn}\tilde V_{ijmn}\frac{Q_M(m,n)}{\omega-
(\epsilon_m+\epsilon_n)+i0}\tilde K_{mnkl}(\omega)
\ee
with
\be
Q_M(m,n)=\left\{ \begin{array}{ll}
1, & \mbox{${min(|{\bf p}_m|, |{\bf p}_n|)>k_{_F}\,\,and\,\,
max(|{\bf p}_m|, |{\bf p}_n|)>k_{_M}}$} \\
0, & \mbox{otherwise.} \end{array} \right.
\ee
The choice of the model-space reaction matrix can avoid the double counting
problem in considering the particle-particle propagations in the ring-diagram
series.\cite{SYK,JKM}

By using the pphh Green function and its RPA equation, the ground-state
energy shift is derived from Eq.(7)
\be
\Delta E_0^{pphh}=\int^1_0d\lambda\sum_m\sum_{ijkl\in
P}Y_m(ij,\lambda)Y^*_m(kl,\lambda)\tilde K_{klij}[E_m(\lambda)],
\ee
where the transition amplitude ${Y_m}$ and eigenvalue ${E_m(\lambda)}$
are solutions of a pphh RPA-type secular equation
\be
\sum_{e,f}[(\epsilon_i+\epsilon_j)\delta_{ij,ef}+(\bar n_i\bar n_j-
n_in_j)\lambda\tilde K_{ijef}(\omega)]Y_m(ef,\lambda)
\nonumber
\ee
\be
=\mu_m(\omega,\lambda)Y_m(ij,\lambda)
\ee
with the self-consistent condition
\be
\omega=\mu_m(\omega,\lambda)\equiv E_m(\lambda)
\ee
and the requirement:
${\sum_{ef}(\bar n_e\bar n_f-n_en_f)|Y_m(ef,\lambda)|^2<0}$. In the above
${n_i}$ is the occupation number which is equal to 1 if ${p_i<k_{_F}}$
and 0 otherwise.
${\bar n_i}$ is defined as ${1-n_i}$. It should be noted that in practice
we have actually include both two-body (the reaction matrix $\tilde K$) and
one-body (self-energy insertion) terms in solving Eq.(11)\cite{SYK}, and
have used the following normalization condition\cite{YK}
\be \sum_{ef}(\bar n_e\bar n_f-n_en_f)|Y_m(ef,\lambda)|^2=\frac{-1}{1-
[\frac{\partial\mu_m(\omega,\lambda)}{\partial\omega}]}.
\ee
As shown later, this brings about the dependence of the ring effect on the
s.p. potential as well as on the nuclear matter density.

Formally speaking, the expressions presented here are rather similar
to those used in the non-relativistic case.
The difference is in the use of relativistic s.p. energies for ${(\epsilon_i)}$
 and in the reaction matrix elements, which are calculated using the
medium-dependent Dirac-spinor  s.p.\ wavefunctions Eq.~(1).

We apply two types of the s.p. spectra, namely a
model-space one and a continuous one. They are defined by
\be
E_p=\left\{ \begin{array}{ll}
\mbox{${\sqrt{(M_{_N}+U_s)^2+{\bf p}^2}+U_v,}$} & \mbox{${if\,\,|{\bf p}|\leq
k_{_M}}$} \\
\mbox{${\sqrt{M_{_N}^2+{\bf p}^2},}$} & \mbox{otherwise}
\end{array} \right.
\ee
where
$k_M$ is finite
for the model space spectrum
(as mentioned, we choose
$k_M=3.2$ fm$^{-1}$) while
$k_M$ is infinite
for the continuous choice.
For each $k_F$, ${U_s}$ and ${U_v}$ are determined from
\be
\tilde u({\bf p}, s)^{\dag}[\gamma_0U_s+U_v]\tilde u({\bf p},s)
=\frac{\tilde M}{\tilde E_p}U_s+U_v
\nonumber
\ee
\be
=2\sum_{h<k_{_F}}<ph|\tilde K(E_p+E_h)|ph>, \hskip 1cm p\leq k_{_M},
\ee
where ${<ph|\tilde K(\omega)|ph>}$ is the relativistic
model-space reaction matrix as obtained from the solution of
Eq.~(8). For the continuous choice, $Q_M$ in Eq.(8) is replaced by the
common Pauli projection operator (i.e. $k_M=k_F$) and a principal value
integration is employed to avoid any singularities.

In our calculations, the first step is to determine for each density
the fields ${U_s}$ and ${U_v}$. This is achieved by
solving Eqs.~(8) and (15) selfconsistently.
The resulting values are listed in Table 1 for both the
model-space and the continuous
s.p. spectrum.
In the next step, the model-space relativistic reaction matrix is calculated
using these self-consistent values and applied in the ring-diagram
formalism.

\section{Results and Discussion}

We have calculated the energy per nucleon in nuclear matter
for various densities applying the relativistic ring method outlined in the
previous section.
For the NN potential we use the relativistic meson-exchange interaction of
the Bonn group~\cite{Pot}, which applies the pseudovector coupling for the $\pi
NN$ vertex. This potential has been used in the Dirac-Brueckner calculations of
Ref.~\cite{Mac,BM90}.
The results of our relativistic ring calculations
are listed in Table~1 and plotted in Fig.~2.
In the figure, the solid line (with
no label) shows the relativistic ring result.
For comparison we also show the non-relativistic ring calculation
using the same potential (solid line with label `nr'). Moreover, relativistic
(DBHF)
and non-relativistic Brueckner-Hartree-Fock results are also shown, again,
using the same potential (dashed curves).

Though,
the difference between the two relativistic calculations is obviously small
(solid and dashed curves with no label), the
effect of the infinite summation of the pphh ring diagrams can still be seen
in the shift of the minimum of the curve towards a smaller density.
{\it Furthermore, the curvature of the relativistic ring curve is
considerably smaller than the one of the DBHF curve, indicating a softer
equation of state for the relativistic ring case.}

As a matter of fact, the shift shows the existence of
an additional density-dependent repulsive effects in the relativistic ring
calculations as compared to the relativistic DBHF calculations. This feature
can be seen more
clearly in Table 2 where we present a $\lambda$-dependence of the average
potential energy as defined by
\be
\Delta E^{pphh}_0=\int^1_0d\lambda\Delta E^{pphh}_0(\lambda).
\ee
As already discussed in previous papers\cite{SYK,JKM},
the ${\lambda}$-dependence characterizes
the relative importance of the higher-order ring diagram contributions.
Thus, the difference between ${\Delta E^{pphh}_0(\lambda=1)}$ and ${\Delta
E^{pphh}_0(\lambda=0)}$ is a direct measure of this contribution. In Table
2, we list ${\Delta E^{pphh}_0}$ for two $\lambda$'s and their density
dependence for the partial-waves $^3S_1$ plus $^3D_1$.
It is clearly seen that
the difference  ${\Delta E^{pphh}_0(\lambda=0.887)-\Delta
E^{pphh}_0(\lambda=0.113)}$ becomes less attractive with increasing
density. This means
a repulsive effect that increases with the density, and therefore
shifts the saturation point.

One can see clearly from Fig.~2 that this ring effect is dramatically reduced
in the relativistic calculations. (Compare the differences between
solid and dashed curves for both the labeled and the unlabeled case.)
This may have something to do with the different momentum dependence of the
relativistic and non-relativistic s.p. spectra. In the relativistic case,
the energy depends approximately linearly on the momentum for large
momentum. In the non-relativistic case, this relation is quadratic for all
momenta. As a result, the $\omega-$dependence of the non-relativistic
eigenvalues $\mu_m(\omega)$ of Eq.(11) would be stronger than that of the
relativistic ones. As seen from Eq.(10), the effect from the infinite
summation of the ring diagrams depends sensitively on the normalization
of the RPA amplitudes $Y_m$ which are normalised using Eq.(13). Therefore,
the stronger the $\omega-$dependence of $\mu_m$, the smaller the $Y_m$
(note $d\mu_m/d\omega<0$), and the larger the ring diagram  repulsive
effect.

For a closer comparison, we list in Table~3 the saturation
properties of nuclear matter as predicted
by the two different relativistic approaches.
Particularly noteworthy is that
in the relativistic ring approach the compression
modulus comes out much smaller than in Dirac-Brueckner calculations.
This implies that
the  equation of state is  predicted softer by the ring approach.

Note that for the relativistic ring-diagram calculations we employ a model
space s.p. spectrum while in DBHF we use a continuous choice. Thus, part of the
observed differences could be attributed to the difference in the choice of the
s.p. potential. To clarify this point, we have also performed a relativistic
ring calculation with a continuous s.p. spectrum (cf.~Table~1).
 It turns out that the results
are, indeed, very similar to the s.p. model space calculation.

We note that in the non-relativistic case\cite{JMK90a}, the differences
between
calculations employing different s.p. potentials are, in general, larger
than in our present relativistic calculations.  This fact can also be
attributed to the different momentum
dependence of the relativistic and non-relativistic s.p. spectra, as we
mentioned earlier.

Comparing the two solid curves in Fig.~2, demonstrates the relativistic
effect in the ring formalism. It is qualitatively of the same kind
as in Brueckner-Hartree-Fock (compare the two dashed curves), but smaller.
This is the consequence of the interplay of the relativistic and ring
effects. The comparison  also tells us that the relativistic effect
is larger and more strongly density-dependent than the ring effect.

Finally, we wish to point out that in contrast to  non-relativistic ring-
diagram calculations, our present calculation is in favor of the NN interaction
with the weaker tensor force. It appears that our results
show support to previous DBHF calculations in the following two aspects:
(1) We need to have a realistic NN interaction, which has a weak
tensor force component. At present, it seems that the Bonn potential is a good
candidate. This observation is also consistent with recent studies in
finite nuclei.\cite{Mac,MMB90,JMSK} (2) Without the consideration of
anti-nucleon contribution, the
infinite summation of the pphh ring diagrams is convergent and its correction
to the DBHF is small. Therefore the DBHF approach may be a sufficient
approximation for most purposes.

\section{Summary and Conclusions}

In this note, we have extended our
 ring-diagram formalism to incorporate  relativistic
effects by substituting the non-relativistic Schroedinger s.p.
wavefunctions by relativistic Dirac-spinor s.p. wavefunctions.
Our results show that in this
relativistic framework, the general feature of the non-relativistic
ring-diagram results are preserved. Thus, the ring effect
shifts the saturation point
to even lower
saturation densities as  compared to Dirac-Brueckner calculations,
in perfect agreement with empirical information.

{\it Furthermore, the equation of state comes out softer in the relativistic
ring approach as compared to Dirac-Brueckner-Hartree-Fock calculations.
Consequently, the compression modulus is predicted smaller
by the relativistic ring formalism.}

As a by-product of our investigation, we find
that the dependence of the nuclear matter energy
on the choice of the s.p. spectrum  is weaker in relativistic calculations.
Mainly because of this, the relativistic pphh ring corrections
to the DBHF nuclear matter calculations have turned out to be relatively small.

\vskip 0.5cm
This work has been supported in part by the U.S. National Science
Foundation under Grant Nos.~PHY-8911040 and PHY-9211607
and the U.S.
Department of Energy (Contract Nos.~DE-FGO2-88ER40388 and
DE-FG02-88ER40425).

\vfill
\eject

\vfill
\eject

\vspace*{-2cm}
\small
{\bf Table\,\,1.}  Energy per nucleon, E/A, and
single-particle potential parameters, $U_s$ and $U_v$, (in units of MeV)
for various densities as obtained in the present
relativistic ring-diagram calculations and in the DBHF
approach of Ref.~\cite{BM90}.
In the case of ring calculations, two types of s.p. potentials
are applied, namely the model-space
and the continuous choices. In the DBHF calculations of Ref.~\cite{BM90}
only the continuous choice was used.
For the definition of ${U_s}$ and ${U_v}$
see Eqs.~(2) and (15).
\begin{center}
\small
\begin{tabular}{cccccccccccc}
\hline\hline
 &\multicolumn{7}{c}{\bf Relativistic Rings}&&\multicolumn{3}{c}{\bf DBHF}\\
 \cline{2-8} \cline{10-12}
$k_F$ &\multicolumn{3}{c}{Model-Space Choice}&&
\multicolumn{3}{c}{Continuous Choice}&&
\multicolumn{3}{c}{Continuous Choice}\\
 \cline{2-4} \cline{6-8} \cline{10-12}
(fm$^{-1}$) & E/A & ${U_s}$ & ${U_v}$ && E/A & ${U_s}$ & ${U_v}$ &&
E/A & $U_s$ & $U_v$ \\ \hline
1.20&-13.97&-305.6&252.4&&-13.75&-288.8&222.0&&-13.44&-288.8&222.0 \\
1.30&-15.00&-344.9&283.5&&-14.66&-331.6&255.0&&-14.86&-331.6&255.0 \\
1.40&-15.29&-386.5&318.1&&-14.80&-374.9&289.8&&-15.59&-374.9&289.8 \\
1.50&-14.62&-431.0&361.3&&-13.83&-416.3&325.7&&-14.88&-416.3&325.7 \\
1.60&-12.35&-474.5&406.9&&-11.18&-459.6&368.6&&-11.96&-459.6&368.6 \\
1.70& -6.77&-505.5&442.2&&-5.64 &-497.2&412.8&&-5.88&-497.2&412.8 \\
1.80&+1.86&-536.5&493.4&&+3.60&-530.4&461.6&&+4.44&-530.4&461.6 \\
1.90&+13.92&-549.5&518.5&&+15.11&-554.8&512.0&&+19.72&-554.8&512.0 \\
2.00&+27.10&-555.9&547.7&&+27.55&-572.4&567.5&&+41.62&-572.4&567.5 \\
2.10&+35.91&-558.1&611.9&&+36.88&-590.2&640.3&&+71.20&-590.2&640.3 \\
\hline\hline
\end{tabular}
\end{center}
\normalsize
\vfill
\eject

{\bf Table\,\,2.} The ${\lambda}$-dependence of the average potential energy
${\Delta E^{pphh}_0(\lambda)}$ (in MeV) as defined in Eq.~(16) in
partial-waves $^3S_1+\:^3\!D_1$ for various densities.
\vskip 1cm
\begin{center}
\begin{tabular}{ccc}
\hline\hline
k$_F$&\multicolumn{2}{c}{${\Delta E^{pphh}_0(\lambda)/A}$} \\ \cline{2-3}
(fm$^{-1}$)&${\lambda=0.113}$&${\lambda=0.887}$ \\ \hline
1.20&-11.30&-20.87 \\
1.30&-12.52&-21.32 \\
1.40&-13.41&-21.41 \\
1.50&-13.94&-21.13 \\
1.60&-14.07&-20.20 \\
1.70&-13.66&-18.68 \\
\hline\hline
\end{tabular}
\end{center}
\vfill
\eject

{\bf Table\,\,3.} Nuclear matter saturation
parameters as obtained in a Dirac-Brueckner-Hartree-Fock (DBHF) and a
 relativistic pphh ring-diagram (RR) calculation. Given are the energy per
nucleon, $E/A$, the Fermi momentum, $k_F^0$, and the compression modulus,
$\kappa$,
 at the saturation point.
\vskip 1cm
\begin{center}
\begin{tabular}{ccc}
\hline\hline
                      &    DBHF      &     RR        \\ \hline
$E/A$ [MeV]           &  -15.60      &   -15.30      \\
$k_{_F}^0$ [fm$^{-1}$]&    1.41      &    1.38       \\
$\kappa$ [MeV]             &    296       &    152      \\
\hline\hline
\end{tabular}
\end{center}
\vfill
\eject

${\underline{\bf{FIGURE\,\,CAPTIONS}}}$
\vskip 2cm
{\bf Figure 1.}
Particle-particle hole-hole ring-diagram series considered in this
study. Each hatched box
represents a relativistic model-space reaction matrix as defined by Eq.(8).
Indices {\it i, j, k, l, m, n} denote model-space single-particle states.

\vskip 1cm
{\bf Figure 2.}
Energy per nucleon in nuclear matter {\it versus}  density expressed in terms
of the Fermi momentum $k_F$. Results from ring-diagram calculations are shown
by solid curves, while Brueckner-Hartree-Fock predictions are represented by
dashed lines. For both approaches, relativistic (no label)
and non-relativistic (label `nr') results are displayed. The shaded box covers
empirical information on nuclear matter saturation ($E/A=-16\pm 1$ MeV,
$k_F=1.35\pm 0.05$ fm$^{-1}$).


\begin{thebibliography}{99}
\bibitem{Pot} In the present work, we employ the Bonn A potential
which is defined in Appendix A, Table A.2, of Ref.~\cite{Mac}.
\end{thebibliography}
\end{document}